\documentclass[english,prl,twocolumn]{revtex4}
\usepackage[utf8]{inputenc}
\usepackage{graphicx}
\usepackage{float}
\usepackage{amssymb}
\usepackage{amsmath}
\usepackage{amsfonts}
\usepackage{dcolumn}
\usepackage{bm}
\usepackage{layout}
\usepackage{color}
\usepackage{array,booktabs,calc}

\usepackage[T1]{fontenc}
\usepackage{array}
\usepackage{makecell}
\newcolumntype{x}[1]{>{\centering\arraybackslash}p{#1}}
\usepackage{tikz}

\makeatletter \@ifundefined{textcolor}{}{
 \definecolor{BLACK}{gray}{0}
 \definecolor{WHITE}{gray}{1}
 \definecolor{RED}{rgb}{1,0,0}
 \definecolor{GREEN}{rgb}{0,1,0}
 \definecolor{BLUE}{rgb}{0,0,1}
 \definecolor{CYAN}{cmyk}{1,0,0,0}
 \definecolor{MAGENTA}{cmyk}{0,1,0,0}
 \definecolor{YELLOW}{cmyk}{0,0,1,0}
 }
\definecolor{orange}{rgb}{1,0.5,0}

\begin{document}

\title{Endohedrally confined hydrogen atom with a moving nucleus }

\author{J. M. Randazzo$^{1}$\thanks{e-mail address: randazzo@cab.cnea.gov.ar} and C. A. Rios$^{1}$}
\affiliation{$^{1}$Divisi\'on de Colisiones At\'omicas, Centro At\'omico Bariloche and CONICET, 8400 San Carlos de Bariloche, R\'\i o Negro,
Argentina.}

\begin{abstract}
We studied the hydrogen atom as a system of two quantum particles in different confinement conditions; a spherical-impenetrable-wall cavity and a fullerene molecule 
cage. The motion is referred to the center of spherical cavities, and the Schr\"{o}dinger equation  solved by means of a Generalized Sturmian Function
expansion in spherical coordinates. The solutions present different properties from 
the ones described by the many models in the literature, where the proton is fixed in space and only the electron is considered as a quantum particle. Our results show that 
the position of the proton (i.e. the center of mas of the H atom) is very sensitive to the confinement condition, and could vary substantially from one state to 
another, from being sharply centered to being localized outside the fullerene molecule. Interchange of the localization 
characteristics between the states when varying the strength of the fullerene cage and mass occurred through crossing phenomena.
\end{abstract}

\maketitle

\bigskip

\section{I Introduction}

Being the simplest atomic system, hydrogen is one of the most extensively studied elements of the periodic table. 
The atom has a relatively simple mathematical description through the analytical solution of the two-body Schrödinger equation, 
which allows a comprehension of its electronic structure, its quantum states, its discrete nature 
of energy levels, and other related properties. From the experimental side, it has been bombarded with electrons 
\cite{ehrhardt_ionization_1969}, protons \cite{shah_experimental_1981,kerby_energy_1995} and photons \cite{beynon_experimental_1965,bergeman_shape_1984}, combined with other 
chemical elements, and more recently \cite{komatsu_encapsulation_2005,lopez-gejo_can_2007} confined in fullerene structures.

The wave functions which theoretically describe these processes are usually obtained from the hydrogen-like Schrödinger equation, by separating the center of mass and relative 
coordinates \cite{cohen-tannoudji_quantum_2006}. The problem maps to a central field problem for a particle of reduced mass $\mu$, and since 
$\mu$ is very similar to 
the electron mass, it is usually interpreted as if the nucleus were fixed at the center of the coordinate system. Actually, the center of mass has also a quantum behavior, 
but it is not generally considered, since the interesting properties such as ionization energies or the excited states structure, only depend on the relative dynamics. 

The same approach is sometimes applied when the atomic confinement is modeled \cite{Amusia_2006,bielinska-waz_spectra_2001,rivelino_configuration_2001,saha_variational_2002}, 
where the proton is generally considered as an infinitely massive particle fixed in space, acting as a Coulomb center for the electrons clamped somewhere in the box 
\cite{neek-amal_ground-state_2007,ferreyra_exact_2013}. In the case where the proton is not centered, the separation of coordinates is 
not possible as discussed by Tanner \cite{tanner_role_1991} and Amore (and Fern\'{a}ndez) \cite{amore_one-dimensional_2010} for the 
harmonic oscillator. The way to deal with the system while keeping the two-body simplicity is considering the electron-wall 
interactions of a spherical box or fullerene molecule through boundary conditions \cite{sech_variational_2011} or central 
potentials \cite{connerade_electron_1999} respectively, as a function of the radial coordinates. The movement of the  nucleus can 
then be considered perturbatively \cite{neek-amal_ground-state_2007}, or in the Born-Oppenheimer approximation, where the energy 
value of the system as a function of its coordinates defines a Potential Energy Surface (PES) through which it moves 
\cite{ferreyra_strong-confinement_1995}.

An alternative way is to consider the system as a three-body problem, consisting of a proton, an electron and a confinement cavity. A first approximation to the ground state 
solution of the hydrogen atom in an infinitely massive spherical box was introduced by F. M. Fernández \cite{fernandez_confined_2010}, who performed a variational calculation with 
a very simple trial function, which results efficient for strong confinement, i.e. $R_{c}$(radius cage)$<1$ a.u. Better results have been found in a more recent calculation 
with a 
generalized Hylleraas basis set of four linear parameters and three exponentials \cite{fernandez_variational_2012}, where mean values of some observables were given for the ground 
state as a function of the confinement radius.

In this work we extend the study of the confined hydrogen atom with a moving nucleus to the case of endohedral confinement. We 
emphasize the analysis to the localization of the proton, identifying the confinement conditions which are (and are not) compatible with the usual atom-centered models. The 
solution of the Schrödinger equation is obtained by means of a Configuration Interaction (CI) approach with Generalized Sturmian Functions, in the coordinates which 
locates the proton and electron from the center of the confinement cage.

The paper is organized as follows: In Sect. II we introduce the driven equations for the system and the mathematical tools we used to solve them. In Sect. III we show 
convergence of the calculations. We present results for the ground state of the Hydrogen atom and describe localization phenomena of the atom for some confinement conditions. 
Finite mass of the confinement molecule and crossing phenomena between the states as a function of the interactions strength and molecular mass is described. Finally, a summary 
and some concluding remarks are given in Sect. IV. Hartree atomic units $\left( \hslash =m_{e}=e=1\right) $ are used throughout this paper.

\bigskip 

\section{II Theory}

\bigskip
We consider the confined hydrogen atom as a system of two quantum particles
inside a spherical cavity of radius $R_{c}$, where the center of the sphere is considered as the
center of the coordinate system. The Hamiltonian can be written as: 
\begin{equation}
H=-\frac{\triangledown _{e}^{2}}{2\mu_{e}}-\frac{\triangledown
_{p}^{2}}{2\mu_{p}}-\frac{1}{m_{c}}\triangledown _{e}.\triangledown _{p}+U_{e}(r_{e})+U_{p}(r_{p})-\frac{\lambda}{r_{ep}}\label{hamiltonian}
\end{equation}
where the sub-index $e$ ($p$) denotes the coordinate and reduced mass associated to
the electron (proton) with the mass $m_{c}$ of the confinement cage and $r_{ep}$ is the electron-proton distance, while $\lambda$ is the strength of the interaction
(i.e. $\lambda=1$ for Hydrogen). We study two cases, the particles inside a spherical box (SB) with impenetrable walls, and a model of endohedral fullerene confinement (FP). 
In the first case, we take $U_{e}=U_{p}=0$ and impose the boundary condition $\Psi (\mathbf{r}_{e},\mathbf{r}_{p})=0$ for $r_{e}$ or $r_{p}$
equal or greater than $R_{c}$.

In the second case, the confinement is associated to a fullerene cage considered to be infinitely massive ($\simeq 720$ times $m_p$). For the electronic potential we use the well 
given by Connerade {\it et al.} \cite{connerade_electron_1999}:
\begin{eqnarray}
U_e(r) = 
\left\{
\begin{array}{cl}
-U_0 < 0 & \hspace{1cm}  r_c \le r \le r_c + \Delta \\ 
0  & \hspace{1cm}  \mathrm{otherwise}   
\end{array}
\right.    
\label{eq:endopotential}
\end{eqnarray}
where $r_c$ and $\Delta$ are the inner radius and the thickness of the shell, respectively. 
We use the values deduced by Xu {\it et al.}~\cite{Xu} ($r_c=5.75$ a.u. and $\Delta=1.89$ a.u.), which are specific
for a $C_{60}$ molecule. On the other hand, the value of $U_0$ is changed in order to explore the general
physics of the system. In a physical picture, this value can actually be changed; for example modifying the number of atoms of the molecule 
and therefore the fullerene structure.

It follows that the complex physics involved in the chemical processes for the $H-C_{60}$ is 
not included in this model. Indeed, the interaction of the proton with the fullerene cage should be different from $U_e$ because of the different physical properties, principally 
its mass, charge and indistinguishablity with other constituent particles of the molecule.
However, since the proton has the opposite electron's charge, it is naturally suggested to consider as a first 
approximation to more elaborate models $U_p(r)=-U_e(r)$. Independently of the veracity of the proposed interaction, the present model will allow us to discover and 
understand some interesting properties of the collective quantum dynamics of two particles with quite different masses.

\subsection{Sturmian expansion}

In order to have a spatial representation of the eigenstates we use the Sturmian basis, which satisfies the equation: 
\begin{equation}
\left[ -\frac{1}{2m _{i}}\nabla _{\mathbf{r}_{i}}^{2}+U_{i}(r_{i})-E_{i}\right] \Psi
_{\nu _{i}}(\mathbf{r}_{i})=-\beta _{\nu _{i}}V(r_{i})\Psi _{\nu _{i}}(%
\mathbf{r}_{i}),  \label{SFEq}
\end{equation}%
together with the physical boundary conditions: 
\begin{eqnarray}
\int |\Psi _{\nu _{i}}(\mathbf{r}_{i})|d\mathbf{r}_{i}^{3} &<&\infty
\label{bep1} \\
\Psi _{\nu _{i}}(\mathbf{r}_{i}) &=&0\hspace{0.25cm}\text{for}\hspace{0.25cm}r_{i}=R_{c},  \label{bep2}
\end{eqnarray}%
where $r_{i}=|\mathbf{r}_{i}|$, $\nu_{i}=\{n_{i},l_{i},m_{i}\}$ and $i=e,p$.

The Sturmian basis results from (\ref{SFEq}) and the conditions (\ref{bep1}) and (\ref{bep2}), by taking the energy $E$ as an externally 
fixed parameter and $\beta_{\nu }$ as the eigenvalue to be determined (here $\nu$ stands for all quantum numbers) \cite{randazzo_generating_2010}. $V$ is any atomic-kind 
potential which depends only on the distance $r_{i}$ and satisfy $V(r_{i})=0$ for $r_{i}>R_{c}$.

Eq. (\ref{SFEq}) is separable in spherical coordinates as usual: \begin{equation}
\Psi_{\nu_{i}}(\mathbf{r}_{i})=\frac{1}{r_{i}}%
S_{n_{i},l_{i}}(r_{i})Y_{l_{i},m_{i}}(\theta_{i},\varphi_{i}),  \label{sep}
\end{equation}%
and we only have to solve for $S_{n_{i},l_{i}}(r_{i})$: 
\begin{equation}
\left[T_{r_{i}}+U_{i}(r_{i})-E_{i}\right] S_{n_{i},l_{i}}(r_{i})=-\beta_{%
\nu_{i}}V(r_{i})S_{n_{i},l_{i}}(r_{i})  \label{ecrad}
\end{equation}
where $T_{r}$ is the radial kinetic energy operator: 
\begin{equation}
T_{r}=-\frac{1}{2m}\frac{d^{2}}{dr^{2}}+\frac{l(l+1)}{2m r^{2}}.
\end{equation}
The boundary condition (\ref{bep1}) ensures the regularity of the Sturmian functions, while the boundary condition (\ref{bep2}) depicts the 
confinement. For the SB case, $R_{c}$ is the radius of
the cage; while for FP it is chosen large enough to affect negligibly the wave functions which are localized by the fullerene 
potential.

As is usual in uncorrelated CI calculations, we use a partial-wave expansion for the Coulomb interaction: 
\begin{equation}
\frac{\lambda}{r_{ep}}=\lambda\sum_{l=0}^{\infty }\frac{r_{<}^{l}}{r_{>}^{l+1}}\frac{4\pi 
}{2l+1}\sum_{m=-l}^{l}Y_{l}^{m}(\hat{\mathbf{r}}_{e})Y_{l}^{m\ast }(\hat{%
\mathbf{r}}_{p}),  \label{eepotential}
\end{equation}%
where $r_{>}$ ($r_{<}$) is the greater (smaller) between $r_{e}$ and $r_{p}$. The eigenfunctions of the
Hamiltonian given by Eq. (\ref{hamiltonian}) can be written as: 
\begin{equation}
\Psi ^{L,M}(\mathbf{r}_{e},\mathbf{r}_{p})=\sum_{\nu }^{N}a_{\nu
}^{L,M}\Phi _{\nu }^{L,M}(\mathbf{r}_{e},\mathbf{r}_{p}),  \label{fdeonda}
\end{equation}
where
\begin{equation}
\Phi_{\nu}^{L,M}(\mathbf{r}_{e},\mathbf{r}_{p})=\frac{S_{n_{e},l_{e}}(r_{e})}{r_{e}}\frac{S_{n_{p},l_{p}}(r_{p})}{r_{p}}
\mathcal{Y}_{l_{e},l_{p}}^{L,M}(\widehat{\mathbf{r}}_{e},\widehat{\mathbf{r}}_{p})
\label{base2e}
\end{equation}
and $\nu =\{l_{e},l_{p},n_{e},n_{p}\}$. Note that we use the spherical bi harmonics $\mathcal{Y}_{l_{e},l_{p}}^{L,M}$,
which are eigenfunctions of the total angular momentum operator and its projection along the $\hat{z}$ axis with
quantum numbers $L$ and $M$ respectively.

By means of the Galerking method and standard algebra packages \cite{lapack} we obtain solutions for the coefficients
$a_{\nu}^{L,M}$ and energies $E$ \cite{randazzo_generating_2010}.

\section{III Results}

\subsection{Convergence properties of the expansion}

The convergence of the wave functions with the number of radial configurations included in the expansion can be analyzed through the ground state energy. Here we consider 
the hydrogen atom ($U_{0}=0$) in a spherical box of radius $R_{c}=10$ a.u., as a function of the radial basis elements per coordinate. Since partial 
wave convergence will be analyzed later, here we study the s-wave ($l_{e}=l_{p}=0$) case. We will use two basis sets. One is the box-based
Sturmians $SF_{b}$ (i.e. $U_{i}=0$ and $V_{i}=1$, $i=e,p$) which depends only on the value of $R_{c}$, the distance at which homogeneous 
conditions are imposed. This is the basis we will use in the next section, where the deep of the potential $U_{0}$ and the coulomb interaction strength $\lambda$ will be varied.

Before we present results, we would like to stress that the Sturmian functions can be defined to efficiently represent a particular state 
\cite{randazzo_generating_2010}. In order to show this, we propose as an example a second basis which gives better results with 
smaller basis elements than the $SF_{b}$. This optimal basis $SF_{o}$ is defined through $U_{i}=0$, $V_{i}=r_{i}^{-1}e^{-\alpha_{i}r_{i}}$ ($i=e,p$), $E_{e}=E_{p}=-0.5$ and  
$\alpha_{p}=2\alpha_{e}=1$ in Eq. (\ref{SFEq}). The potentials and parameters we have just defined were chosen based in our previous experience 
\cite{randazzo_generating_2010} and performing few variational iterations. A deeper optimization can also be performed in all parameters and functional space of 
potentials $V_{i}$.

In Table \ref{evsnr} we show the ground state energies of the s-wave ($l_{e}=l_{p}=0$) confined H atom as a function of the
number of radial states per coordinate. We clearly see that the optimized basis is much more efficient than the
unoptimized one. However, the disadvantage of optimization is that it has to be performed for each individual state in an iterative procedure. Here 
we reach, with the $SF_{o}$, convergence of the radial expansion for the ground state of the system until the 6th decimal figure by using 
25 radial functions per coordinate, while for the same size the $SF_{b}$ basis gives only up to the first one.

\begin{center}
\begin{table}
{\centering \begin{tabular}{|c|cc|}
\hline  & \multicolumn{2}{c|}{Ground state energy} \\ \hline\hline
\hspace{0.5cm}  $n_{max}$ \hspace{0.5cm} & \hspace{0.cm} $SF_{b}$ \hspace{0.5cm}& \hspace{0.5cm}$SF_{o}$\\
\hline
5 &    -0.2905806409 & \hspace{0.5cm}-0.4470251209\\
10&    -0.3729047660 & \hspace{0.5cm}-0.4663337297\\  
15&    -0.4130974121 & \hspace{0.5cm}-0.4672219778\\
20&    -0.4353883725 & \hspace{0.5cm}-0.4672434215\\
25&    -0.4485486681 & \hspace{0.5cm}-0.4672446242\\
30&    -0.4565116034 & \hspace{0.5cm}-0.4672448546\\
35&    -0.4613104686 & \hspace{0.5cm}-0.4672448585\\
40&    -0.4641220711 & \hspace{0.5cm}-0.4672449074\\
45&    -0.4656910723 & \hspace{0.5cm}-0.4672448904\\
50&    -0.4665127698 & \hspace{0.5cm}-0.4672449113\\   \hline
    \end{tabular}
  }
  \centering 
  \caption{Convergence analysis for the $l_{e}=l_{p}=0$ ground state of the confined hydrogen atom as a function of
radial functions per particle $n_{max}$ with the box based ($SF_{b}$) and ``sophisticated'' ($SF_{o}$) radial Sturmians.}
  \label{evsnr}
\end{table}
\end{center}

Deeper values of the energy can be obtained by adding angular terms. This is done through the bi spherical
harmonics, in the same fashion as for the $SF_{b}$ and $SF_{o}$ basis. The expected energy for the ground state will be
close to $-0.5$, which is a little bit higher due to the finite size of the basis and the confinement effects. If we were dealing with an H atom modeled as a single neutral 
particle in a box, the amount of the total energy associated to the confinement effect would be $\pi^{2}/2R_{c}^{2}(m_{e}+m_{p})\simeq 2.7 \times 10^{-5}$, i.e., the exact
ground state energy of the system. However, as we will see, the true state corresponding to a two particle wave function is very different in
shape if we look at the center of mass distribution, which almost equals the proton's distribution. That is why we can't estimate a priori
the confinement energy of the composed system.

\begin{center}
\begin{table}
{\centering \begin{tabular}{|c|cc|}
\hline  & \multicolumn{2}{c|}{Ground state energy} \\ \hline\hline
\hspace{0.5cm}  $l_{max}$ \hspace{0.5cm} & \hspace{0.cm} $SF_{b}$ \hspace{0.5cm}& \hspace{0.5cm}$SF_{o}$\\
\hline
0	&-0.4353883725	& \hspace{0.5cm}	-0.4672434215\\
1	&-0.4748432442	& \hspace{0.5cm}	-0.4878532032\\
2	&-0.4843652580	& \hspace{0.5cm}	-0.4922149011\\
3	&-0.4884665735	& \hspace{0.5cm}	-0.4934791747\\
4	&-0.4907433951	& \hspace{0.5cm}	-0.4939416245\\
5	&-0.4921976351	& \hspace{0.5cm}	-0.4941458680\\
6	&-0.4932104410	& \hspace{0.5cm}	-0.4942489924\\
7	&-0.4939553268	& \hspace{0.5cm}	-0.4943062823\\\hline
    \end{tabular}
  }
  \centering 
  \caption{Convergence analysis for ground state of the confined hydrogen atom as a function of
 the maximum number of partial waves ($l_{max}$) added when using $20$ radial functions per coordinate, with the box based ($SF_{b}$) and 
``sophisticated'' ($SF_{o}$) radial Sturmians.}
  \label{evsnl}
\end{table}
\end{center}

In Table \ref{evsnl} we show the energy values of the ground state energy of the system as a function of the maximum
number of partial waves used, when the number of radial functions per coordinate is 20 (note that the values for
$l_{max}=0$ are the same as the value for $n_{max}=20$). We clearly see an improvement in the energy value when
increasing $l_{max}$.

As we are not introducing the relative coordinate $r_{ep}$ in our expansion, the convergence of the energy is much less efficient than in the case of the generalized Hylleraas 
series, since with fewer basis functions, the Hylleraas expansion gives better results. Moreover, if one considered the spherical coordinates we use, the Kato cusp conditions 
\cite{Kato0} would be found in the angular coordinates. This sharp behavior cannot be reached with spherical harmonics. However, the CI calculation with a large number of 
configurations is more complete, in the sense that a great variety of distributions can be described with it. In fact, the eigenvalue calculation which gives the last element of 
Table (\ref{evsnl}) corresponds to an expansion with $20\times20\times (7+1)=3200$ basis elements, which gives rise to the same number of eigenstates. In our case we do not know 
the distribution we are going to find, so the basis must be general and not optimized for hydrogenic states.

\subsection{\protect\bigskip Ground State 1S: spatial distribution}

Let us now study how the confinement influences the wave functions to determine the particle's spatial distribution. 
Starting with the spherical box considered in the previous section, we vary the interaction between the particles through the $\lambda$ 
parameter (shown in Eq. (\ref{hamiltonian})) and the strength $U_{0}$ that describes the interaction with the fullerene cage, taking into 
account only s-wave partial wave terms. The results are summarized in Fig. (\ref{estados}):


\begin{center}
\begin{figure}
\includegraphics[angle=0,width=8cm]{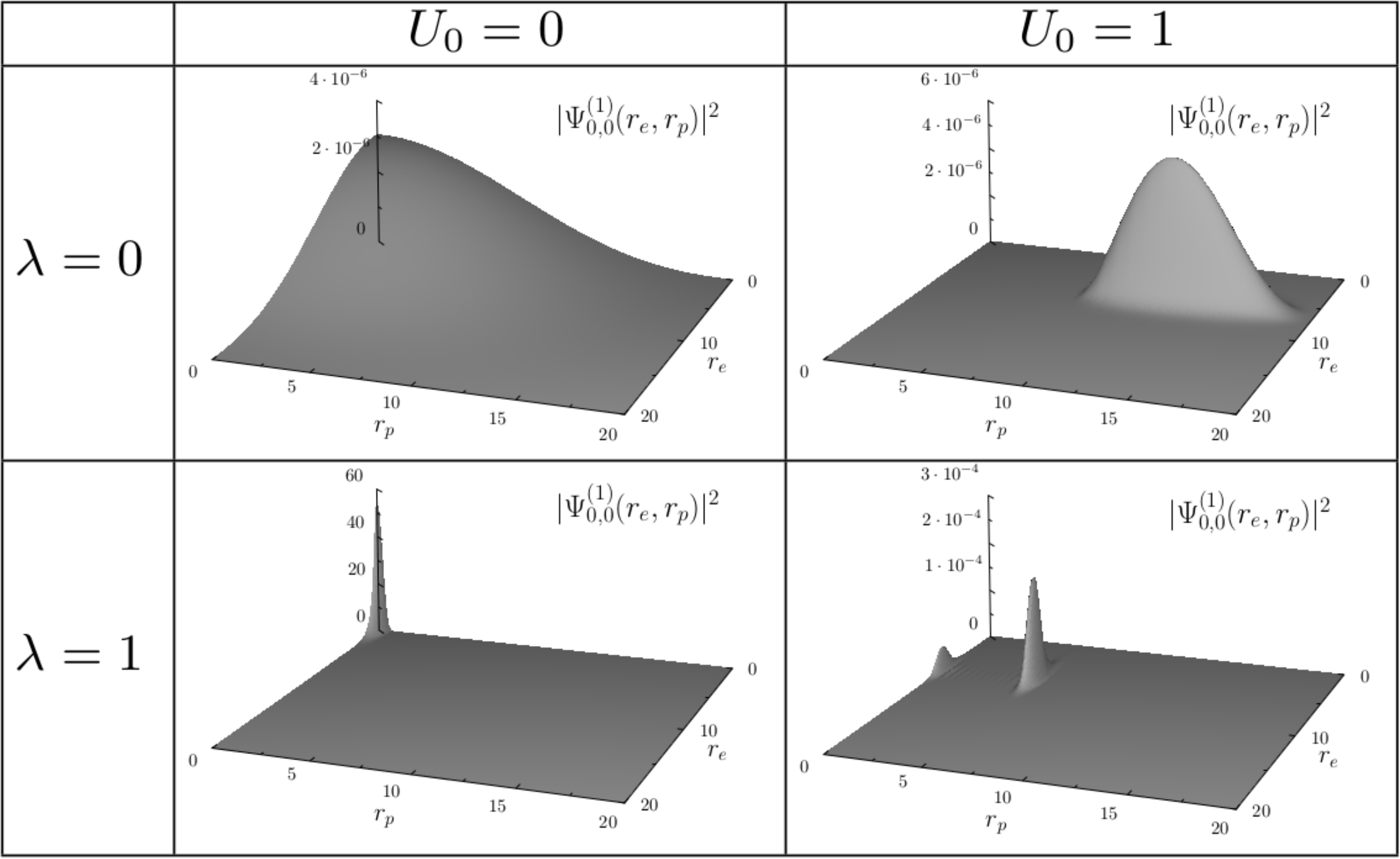}\caption{Spatial distributions of the ground states of the e-p system under different confinement conditions which are simulated by 
varying the strength of the interactions through the values of $\lambda$ and $U_{0}$ (see text).}
\label{estados}
\end{figure}
\end{center}

The first solution is obtained by setting $\lambda=U_{0}=0$ (top-left), and corresponds to box based behavior in each coordinate. The Hamiltonian is uncorrelated in the 
variables, and the solution is exact and can be described by a single product of Sturmian functions. When the $C_{60}$ potential model is turned on to $U_{0}=1$  
($\lambda=0$) (top-right), the electron moves to the well while the proton, which sees a repulsive barrier, stays in the external region. It is still contained by the walls at 
$R_{c}$, but if an infinite radial domain were used, its spatial distribution would be that of an unbounded particle. If the Coulomb potential is turned 
on in absence of the fullerene well, ($\lambda=1$ and $U_{0}=0$), both particles get close to the center of the sphere, as it is shown in the bottom-left part of Fig. 
(\ref{estados}). This appears to be a strange behavior, since the global $H$ system is neutral, and it would be expected the distribution of a particle of mass $m_{e}+m_{p}$ 
inside a spherical well for the center of mass of the $H$ atom, which is practically coincident with the proton position. Instead, we find a very sharp wave function, product of 
the interaction of the complex two-particle system with the boundary. In order to understand this behavior, we have to take into account that the proton is much heavier than the 
electron, and its confinement is less expensive from the energetic standpoint.

Since the ground state lacks nodal curves \cite{mitnik_endohedrally_2008} and must be null at the surface of the confining sphere, we supposed the wave function would be a 
decreasing function of the electronic coordinate, having its maximum at the center of the sphere. This is true for small $R_{c}$, where confinement energy becomes more important 
than Coulomb interaction \cite{Aquino_PLA(2003)}. Such distribution would act as a potential energy surface ($PES$) for the proton, giving a distribution of it centered at its 
minimum (the maximum of the electronic distribution) with a very small width because of its mass. 
Now we can argue that the electronic distribution of this particular state would be the one that adjusts to a proton located at the center of the
coordinates \cite{bielinska-waz_spectra_2001,sech_variational_2011,tanner_role_1991,Amusia_2006}, as many models established. We have increased the value of the cage radius up to 
$R_{c}\simeq 100$ $a.$ $u.$ and found the same centered distribution.

\begin{center}
\begin{figure}
\includegraphics[angle=0,width=8cm]{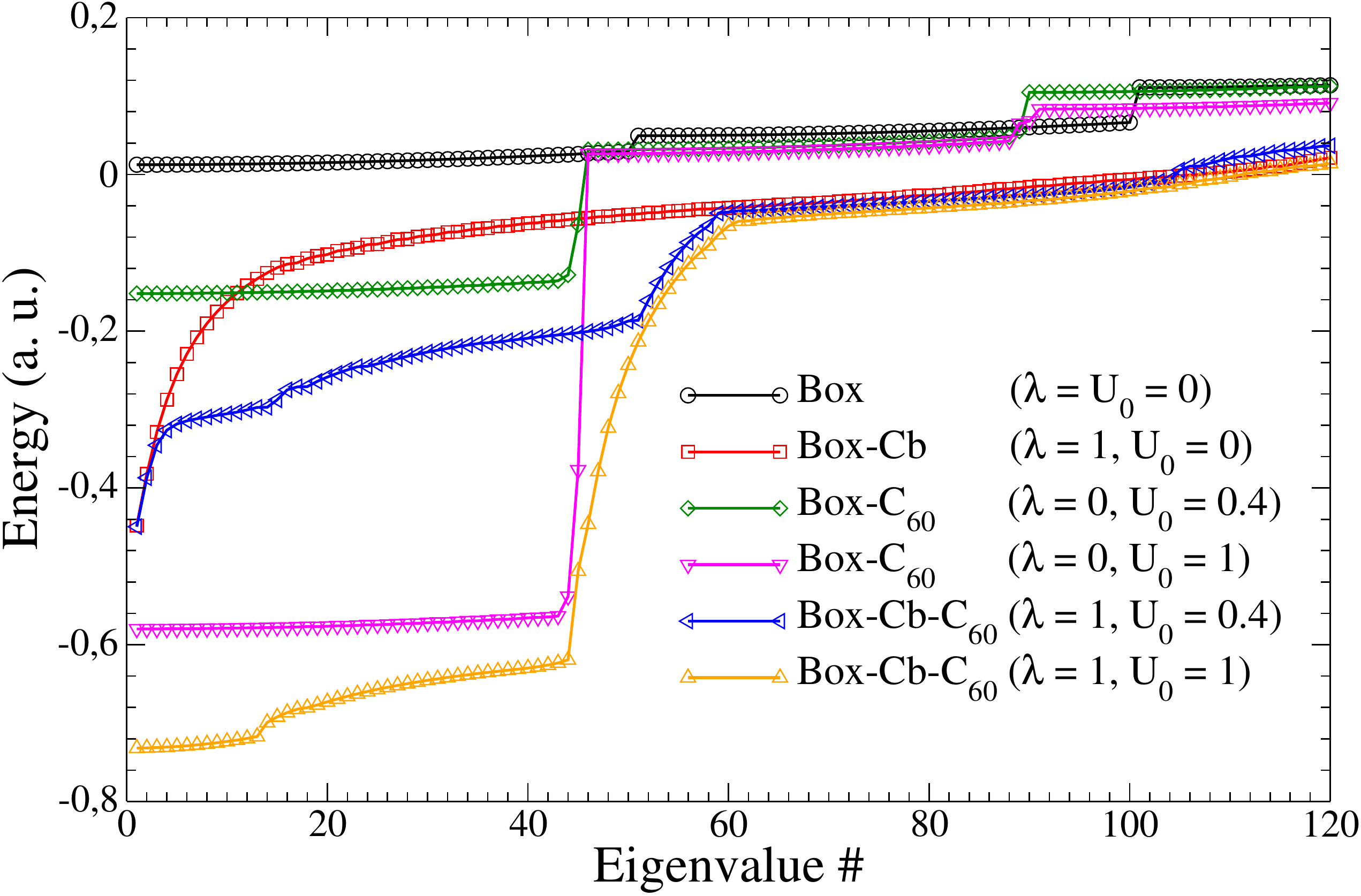}
\caption{Energy value vs. its order of appearance for the confined e-p system under various situations. Box conditions are imposed at 
$R_{c}=20$ and several parameters describe the interactions. (Black) circles: $\lambda=U_{0}=0$; (red) squares: 
$\lambda=1$ and $U_{0}=0$; (green) diamonds: $\lambda=0$ and $U_{0}=1$; (magenta) down triangles: $\lambda=0$ and $U_{0}=1$; (blue) up triangles: $\lambda=1$ and $U_{0}=0.4$; 
(orange) left triangles: $\lambda=1$ and $U_{0}=1$.}
\label{eigen}
\end{figure}
\end{center}

The plot of the bottom-left part of Fig. (\ref{estados}) corresponds to the ground state of two interacting coulomb particles in presence of the $C_{60}$ 
potential ($\lambda=1$ and $U_{0}=1$). In this case the wave function describes the electron in the well while the proton, attracted by the electron and repelled by 
the potential, is located in the internal part of the fullerene cage. This result shows a significant difference with the models of endohedral 
confinement with the centered atom.

\subsection{Behavior of the spectrum}

To deepen the understanding of the results, let us analyze the energy spectrum for different values of $U_{0}$, with and without 
Coulomb interaction. Several of these statements are conclusions drawn from the analysis of both spectra graphs as several charts of wave functions for different conditions, 
analogous to those shown in Fig. (\ref{estados}).

The energies are shown in Fig. (\ref{eigen}), where we plot them in increasing order for the s-wave model ($l_{e}=l_{p}=0$) in a calculation with $50$ radial functions 
per coordinate, making a total of $2500$ states. Here we use box based Sturmians  ($U_{i}=0$ and $V_{i}=1$, $i=1,2$ and $R_{C}=20$ in Ec. ( \ref{ecrad})), so that all uncorrelated 
calculations ($\lambda=0$) are exact solutions. These are represented with black circles, and their values, obtained from the calculation, can also be determined by the 
expression: 
\begin{equation}
E=E_{p}+E_{e}=(\pi n_{e})^{2}/2R_{c}^{2}m_{e}+(\pi n_{p})^{2}/2R_{c}^{2}m_{p}, 
\label{EnerPE}
\end{equation}
the grounds state having energy equal to $E=0.012343$ a.u. and corresponds to $n_{e}=n_{p}=1$.

The first fifty states correspond to the case $n_{e}=1$ and $n_{p}=1,2,...,50$, the small difference between them being a consequence of the mass value of the proton. A jump 
appears at the state number $51$, which corresponds to the case $(n_{e}=2,n_{p}=1)$, followed by the $(n_{e}=2,n_{p}=2,3,...,50)$ states, and so on. We must clarify that the 
jump between the $n_{e}=1$ and $n_{e}=2$ would be unnoticed if a minimum number of proton states were included  such that $n_{p}\ge \sqrt{4m_{p}/m_{e}}\simeq 86$, which follows 
from Eq. (\ref{EnerPE}). 

A similar structure appears when correlation or other values of $U_{0}$ are considered. Coulomb interaction (red squares) with $U_{0}=0$  
changes the nodal structure, laying the $(n_{e}=2,n_{p}=1)$ state of energy $-0.118891$ a.u. in the 15th place (instead of the 51th) and moving down the whole 
set of values, the jump being now unobserved. For higher energies the Coulomb interaction becomes relatively less important than the confinement, and the slope of the curve 
(actually, a quadratic behavior) tends to that of the uncorrelated case. Ground state energy is $-0.447973$ a.u., which is close to the ground state of the free hydrogen system.

With $\lambda=0$ and $U_{0}=0.4$ (dark green triangle down), the first energy values are lower than for the $U_{0}=0$ case. 
On the one hand, this behavior is in accordance with the attractive effect of the $C_{60}$ potential on the electron. On the other hand, it is repulsive for the proton 
and the effect should be the contrary. As we mentioned, the proton localization, in this case outside the repulsive barrier, is less 
expensive from the energetic point of view, while the bound energy of the electron in the well dominates.
The $n_{e}=1\to 2$ ``jump'' occurs now between the 44th and 46th places, with the 45th eigenvalue in a transition region, and corresponds to electron states which are bound for 
$n_{e}=1$ and unbound for $n_{e}=2$. The same happens with $U_{0}=1$ (magenta diamonds), with a bigger jump, since the effect of the potential is higher, still having only 
one electron bound state.

When $\lambda=1$ and $U_{0}=0.4$ (blue triangle left), it corresponds to the confined $H@ C_{60}$ model. The ordered energies 
present a structure which seems to be an admixture of the ones discussed above. We see that the energies of the first and second 
states are practically unchanged from the 
$U_{0}=0$ case. This is because the wave function is centered and without overlapping the well. Then they follow states with the 
electron bounded to the cage and the proton 
inside and outside the cage. States 50th to 59th, have mostly distributions in which the electron-proton pair is bounded and ``glued'' to the surface from the inner 
side, while in the 60th both particles are outside the fullerene. For $\lambda=U_{0}=1$ (orange triangle up) the structure is similar, but with a wither Coulomb structure (from 
state 44th to 60th) and lower ground state. In this case the ground state energy is lower than $-0.5$ a.u., corresponding to the situation where the electron is in the well (lower 
right of Fig. (\ref{estados})). There exists, however, a state with energy $-0.446197$ a.u. which has almost the same probability distribution as the ground state for $U_{0}=0$. 
This state is in the 46th place, and consequently has a complex (hidden) nodal structure which is not present in the 
ground state for $U_{0}=0$.

The interchange of properties in the probabilities between different eigenstates when a parameter of the equation is changed, is related to what is called \textit{crossings} of 
energies, and will be discussed in the next section. 

\subsection{\protect\bigskip Crossing phenomena}

\bigskip 

The behavior of the electronic energy levels as a 
function of the magnitude of the fullerene potential has been deeply studied for a confined Helium model \cite{mitnik_endohedrally_2008}. In 
particular, the phenomenon of ``mirror collapse'', where the wave functions interchange their characteristic probability 
distribution while keeping its position in the ordered list of eigenvalues and conserving its nodal structure. Without making 
an exhaustive description of the behavior of the states around each crossing, we want to show that the same kind of phenomena presents here the endohedrally confined hydrogen atom 
with a moving nucleus. This time we also show what happens when we vary the mass of the fullerene cage while keeping the 
magnitude of the $C_{60}$ potential fixed.

\begin{center}
\begin{figure}
\includegraphics[angle=0,width=8cm]{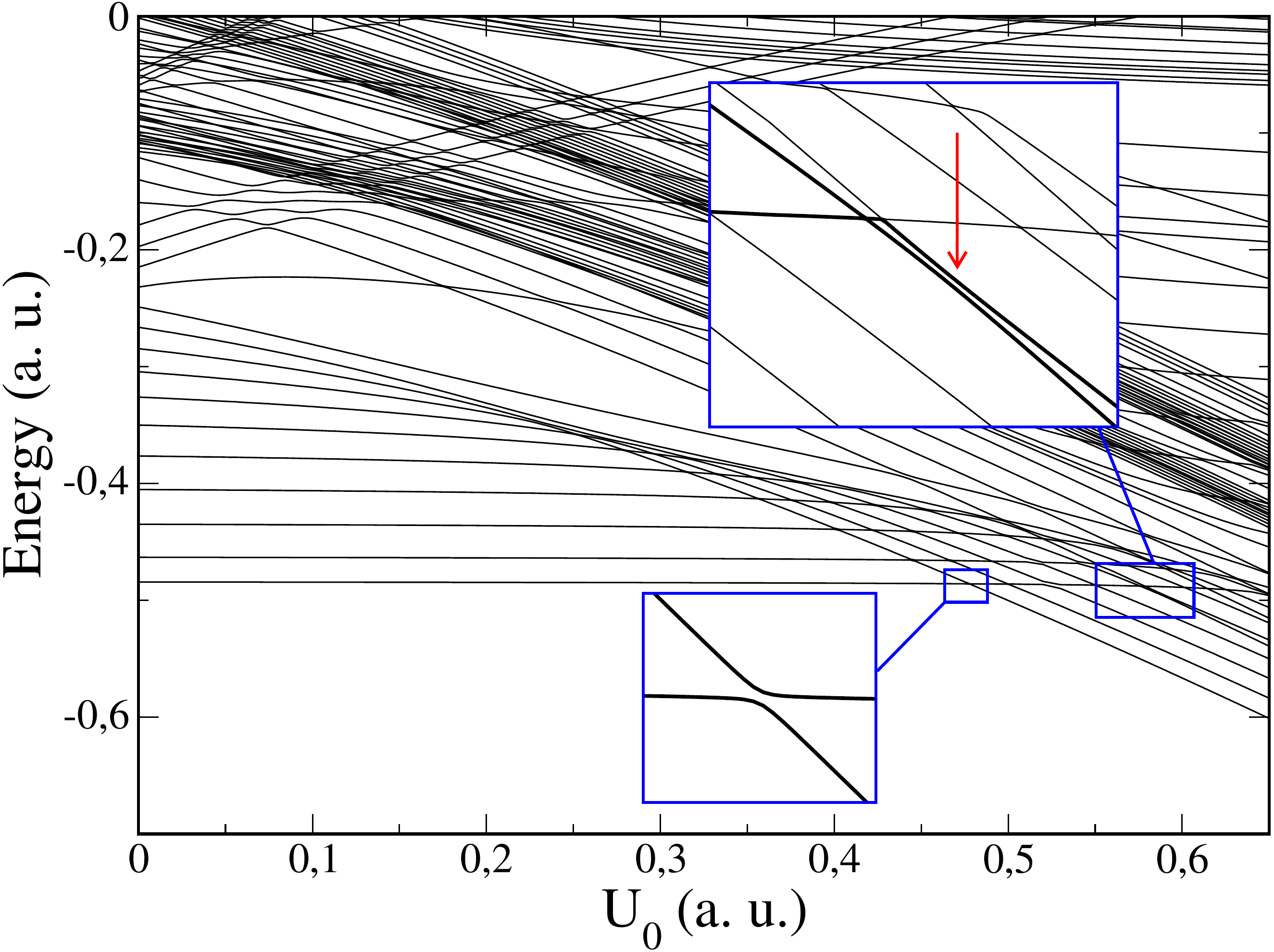}
\caption{Evolution of the negative energy levels of the confined hydrogen atom as a function of $U_{0}$.}
\label{EvsU0}
\end{figure}
\end{center}

Fig. (\ref{EvsU0}) shows the evolution of the negative energy levels as a function of $U_{0}$. We see that the ground state energy is 
not equal to $-0.5$, but higher. In this case we use $20$ $SF_{b}$ per radial coordinate and $l_{max}=2$. This is because of the 
confinement and also the finite size of the partial wave and radial expansions. Although the energy values are not too 
precise with these basis dimensions, the results are representative enough for the understanding of the phenomena we want to 
describe.

\begin{center}
\begin{figure}
\includegraphics[angle=-90,width=8cm]{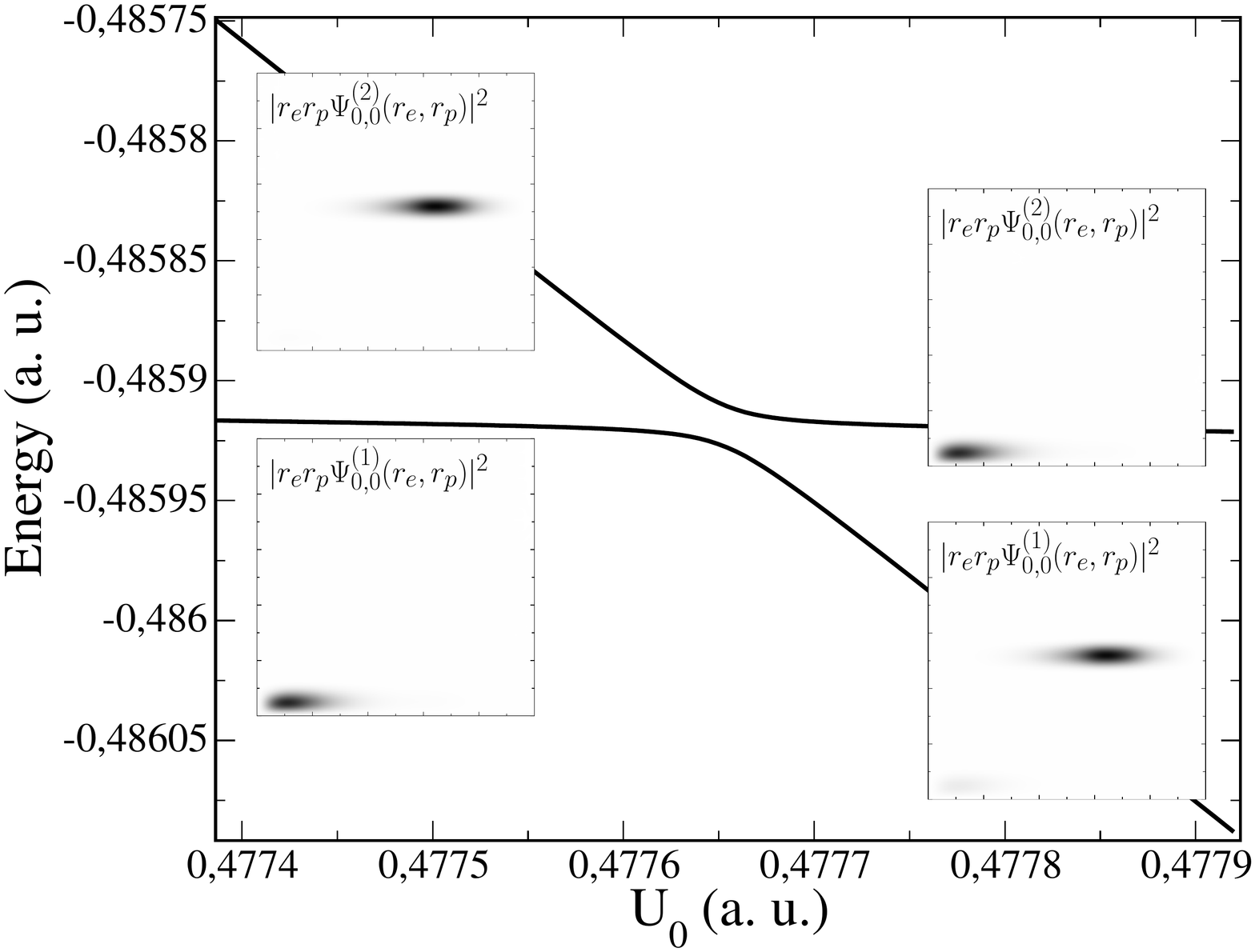}
\caption{Mirror collapse between the first two states of the confined H system for the evolution of the energy values as a function of the 
potential $U_{0}$. Inner boxes show the spatial distribution of the two states at the left and right sides of the crossing, where it can be 
seen that the spatial distribution of the particles becomes interchanged.}
\label{crossing}
\end{figure}
\end{center}

We can observe an avoided crossing between the first two energy levels as a function of $U_{0}$, which occurs around $U_{0}=0.47$. 
At the inner box of Fig. (\ref{EvsU0}), we see a magnification of this crossing, and at Fig. (\ref{crossing}) the density 
plots of the two-dimensional probability distributions for both states at the left and right of the crossing, where it clearly shows that the interchange of the probabilities 
occurs. In this case, the ground state behavior corresponds to the atom centered at the cage in a typical ``free'' hydrogen distribution, 
while in the first exited state the bounded pair is located close to the surface, with the proton and the electron within and outside the $C_{60}$ potential.

To understand this behavior, we have to note that for small values of $U_{0}$ the ground state has practically no overlap with 
the fullerene well, so its energy does not depend on $U_{0}$. The first excited state correspond to the case where the electron 
lies in the well and the proton is around it. Again, localization of the proton is less expensive from the kinetic energy point of 
view, and it tends to locate as close to the electron as possible, without overlapping the repulsive region. The energy of the 
first exited state linearly decreases with $U_{0}$ at $0.45$, and at $U_{0}=0.47$ it encounters the ground state energy. At the same time, the 
ground state pair starts to move to the well gradually since its configuration there has the same energy as at the center of the 
cage. As the states become quasi degenerate, their nodal structure is maintained, with the ground state never having nodal surfaces. 
The crossing is ``avoided'', i. e. the states never have exactly the same energy value for a fixed $U_{0}$.

\begin{center}
\begin{figure}
\includegraphics[angle=-0,width=8cm]{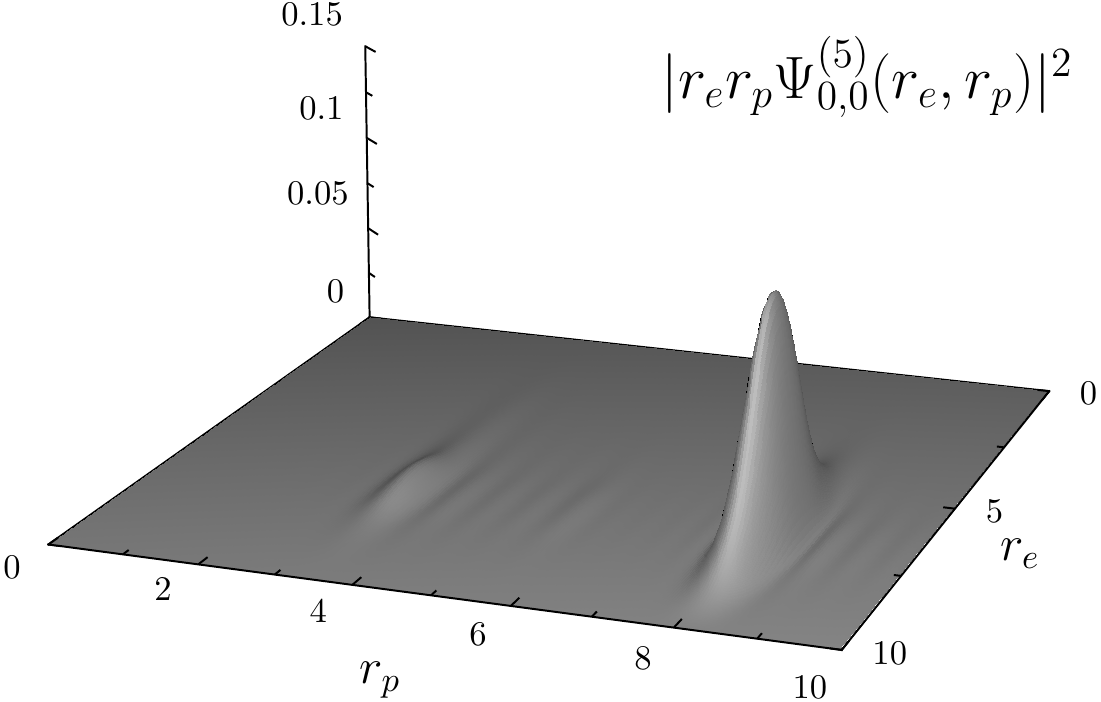}
\caption{Three dimensional plot for the 4th excited state of the confined hydrogen atom, of energy $-0.573$ for $U_{0}=0.575$. Distribution shows the proton localized outside the 
fullerene structure.}
\label{outside}
\end{figure}
\end{center}

A striking distribution appears for $U_{0}\ne 0$, and corresponds to a bound state where the proton is mostly outside the fullerene. The distribution mixes 
through the crossing structure of the Fig. (\ref{EvsU0}) and is amplified in the upper inner box. It corresponds to the 4th and 5th excited states, having an energy 
close to $-0.573$ (lower than the hydrogen bound state). In Fig. (\ref{outside}), the spatial distribution for the 4th state is shown and is related to a bound state of the system 
in 
which the proton distribution is \textit{outside} the fullerene cage. Crossing appears close to $U_{0}=0.575$ and is shown in Fig. (\ref{crossing}). 
Note that, asymptotically, the field in which the proton moves is Coulombic. As it is known for potentials with asymptotic Coulomb tails \cite{PhysRevD.33.1745}, there exists an 
infinite number of bound states. We estimate that there exists an infinite number of states whit these characteristics for the proton distribution (i. e. located outside the 
fullerene cage).

\begin{center}
\begin{figure}
\includegraphics[angle=-90,width=8cm]{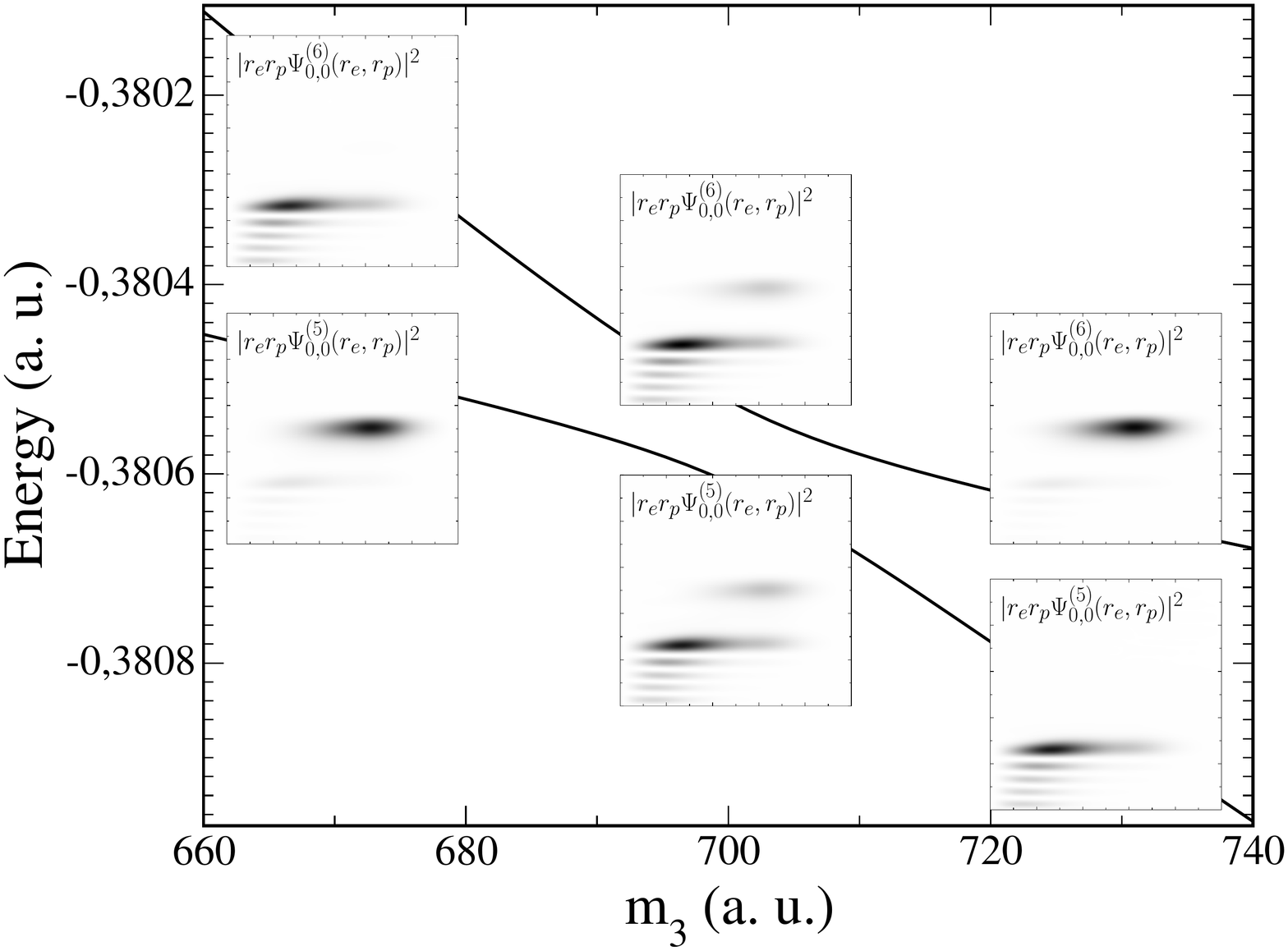}
\caption{Mirror collapse between the two first states of the confined H system for the evolution of the energy values as a function of the 
fullerene mass $m_{3}$. Their properties become interchanged when the $C_{60}$ mass vary from $660$ to $740$ electron 
mass.}
\label{crossing2}
\end{figure}
\end{center}

Finally, we vary the mass of the fullerene molecule in search of crossings, from (nonphysical) values $\simeq 1$ $a. u.$ to $1.44\times 10^{6}$ $a. u.$ (the order of magnitude of 
the exact value for fullerene molecule). We found that all the crossings occur at mass values which are small compared to the mass of the $C_{60}$ fullerene or other species 
$C_{n}$. However, we found a situation we think is interesting to describe, and which could be present in confinement of heavier atoms or 
molecules. Fig. (\ref{crossing2}) shows the crossing between the 5th and 6th energy levels when $m_{3}$ vary from $660$ to $770$ $m_e$.

In this particular case, the localization of the particles (shown in the inner boxes) becomes also interchanged and corresponds to 
the case where the atom localize inside and outside the $C_{60}$ potential. The distribution of both states represents very different situations for the proton distribution, one 
where the proton has an oscillating inner distribution in the region $r_{p}\lesssim 4$, and the other glued to the fullerene surface from the inner side ($4 \lesssim r_{p}\lesssim 
6$).

\section{IV Conclusions}

We have studied the confinement of the hydrogen atom in a spherical cavity through a quantum mechanical model which included both,
the electron and the proton dynamics. We have considered the case of a spherical well of impenetrable walls and a fullerene $C_{60}$ which was
modeled as a spherical barrier, attractive for the electron and repulsive for the proton. The dynamic was driven by the solutions of the three-body-like Schrödinger 
equation, where the motion of the particles is described in the coordinates which locate them with respect to the center of the spherical cavity. The solution 
was obtained through the Generalized Sturmian Functions method.

The wave functions showed a behavior that substantially differs from the many models found in the literature, where the proton is fixed 
at the center of the coordinate system, or moved perturbatively in the Born-Oppenheimer approximation. In the moving nucleus case, the states showed a different proton 
distribution for each energy level, being the bound state with the proton located outside the fullerene, one of the more curious situations which manifested the complexity of the 
system. For a fullerene potential which was deep enough to keep the electron bounded at least in one state, we can argue that the asymptotic interaction for the proton 
corresponded to an attractive Coulombic tail, which supported infinitely many bound states, in particular Rydberg states.

The spectrum of the system in the moving nucleus case was much more dense than the given for the fixed nucleus. The ``new'' levels were associated to the dynamics of the proton, 
and can be understood by considering the distance between the energies for a confined particle decreasing with the inverse of the mass.

Apart from the complexity of the moving nucleus system, we found one match with the fixed nucleus model for the ground state of weak fullerene wells. In that case,
the proton located at the center of the molecule, and the same happened for the rigid sphere. The corresponding distribution of the center of mass of 
the atom was very different to the box ground state of a neutral particle. It reflected how different the dynamics of a simplified model could be from a composed system.

Finally, we found crossing phenomena for the spectrum as a function of the parameters of the equation. By varying the magnitude of the fullerene well, we found many 
avoided crossings, while by changing the fullerene mass, we found only one. These are the preliminary results of a more exhaustive searching of crossings which will include the 
variation of radius of the fullerene well. This change of radius would actually happen if vibrational modes of the carbon structure were considered. The results would 
be part of an oncoming publication.

\section{Acknowledgments}

This work has been supported by ANPCyT (PICT-2014-1400) and CONICET (Argentina). J. M. Randazzo thanks
the support of Universidad Nacional de Cuyo through grant 06/624.

\bibliographystyle{apsrev}
\bibliography{BIBLIOGRAFIA}
\ 

\end{document}